\documentclass{svmult}

\usepackage{amsmath}
\usepackage{color}
\usepackage{cite}
\usepackage{amsfonts}

\usepackage{mathptmx}       
\usepackage{helvet}         
\usepackage{courier}       
\usepackage{type1cm}

\newcommand\bZ{\mathbb Z}
\newcommand\bT{\mathbb T}

\newcommand \tr {\operatorname{tr}}

\newcommand{\rmi}{\mathrm{i}}
\newcommand{\rme}{\mathrm{e}}
\newcommand{\rmd}{\mathrm{d}}

\begin{document}

\motto[7.5cm]{We dedicate this paper to the memory of George Sudarshan, with gratitude for the many years of friendship and for sharing with us his great insights in physics.}

\title*{Topological order, mixed states  and open systems}
    \author{Manuel Asorey, Paolo Facchi \and Giuseppe Marmo }
\institute{
Manuel Asorey \at
Centro de Astropart\'{\i}culas y F\'{\i}sica de Altas Energ\'{\i}as, Departamento de F\'{\i}sica Te\'orica,\\
Universidad de Zaragoza, E-50009 Zaragoza, Spain
\\ \email{asorey@unizar.es}
\and Paolo Facchi \at
Dipartimento di Fisica and MECENAS, Universit\`a di Bari, I-70126  Bari, Italy \\
INFN, Sezione di Bari, I-70126 Bari, Italy
 \\  \email{paolo.facchi@ba.infn.it}
\and Giuseppe Marmo \at   Dipartimento di Fisica, Universit\`a di Napoli Federico II,
I-80126 Napoli, Italy 
\\ \email{marmo@na.infn.it}
}

\maketitle

\abstract{The role  of mixed states in topological quantum matter is less known than
that of pure quantum states. Generalisations of topological phases appearing in
pure states had received only quite recently attention in the literature. In particular, it is still unclear whether 
 the generalisation of the
Aharonov-Anandan phase for mixed states  due to  Uhlmann plays any physical  role 
in the  behaviour of the quantum systems.
 We analyse from a general viewpoint  topological phases  of mixed states and 
 the robustness of their  invariance. In particular,  we analyse the role of these phases in
 the behaviour of systems with a periodic symmetry and their evolution under the influence of
 an environment preserving its crystalline symmetries.}

\section{Introduction}
\label{sec:1}

Although the existence of topological phases of quantum theories is known since a  long time ago (see e.g.~\cite{phases} for a review and  references therein), topological aspects of quantum matter have been   intensively exploited only in recent years.
Topological phases are characterised in terms of topological invariants and some discrete symmetries such as time reversal invariance.   The robustness of the corresponding effects under perturbations follows from the topological nature of the phenomena, especially under two kinds of disorder perturbations:  either generated by impurities or by 
small deformations of the material.
Most of the studies where formulated in terms of pure quantum states. In this paper we analyse the relevance of such topological invariants for mixed states of closed and open quantum systems, corresponding for example to electrons in crystalline solids either in   isolated conditions or  under the effects of interactions with the environment.

\section{Quantum states, principal fibre bundles and geometrical phases}

According to Dirac, the simplest way to take into account interference phenomena is to associate with every quantum system a Hilbert space $\mathcal{H}$. The evolution is ruled by the linear Schr\"odinger equation and solutions may be ``superposed''. 
Then the evolution of a quantum system can be considered as a parallel transport of unitary operators acting on a bundle of Hilbert spaces along the time axis with respect to a generalised connection associated with the Hamiltonian operator~\cite{ACP}. If the Hamiltonian has a smooth dependence on a family of parameters with a cyclic symmetry, an adiabatic evolution of the system can develop a cyclic evolution of quantum states. Moreover, the final state can have a phase different from that of the initial state. The difference between these phases is known as Berry phase~\cite{Berry}. Such a phase difference has one component which is dynamical, depending on the Hamiltonian, and another one which has a purely geometric origin. This component goes often under the name of Aharonov-Anandan phase~\cite{AA}.

\subsection{A fibre bundle description of the Aharonov-Anandan phase}

To avoid technicalities we shall restrict our considerations to finite dimensional quantum systems. The Hilbert space is then $\mathcal{H}=\mathbb{C}^{N}$, and we shall denote by $\mathcal{H}_0=\mathbb{C}_0^{N} = \mathbb{C}^{N}\setminus \{0\}$ the space deprived of the zero vector. The probabilistic interpretation of quantum mechanics requires that pure states are rays of $\mathbb{C}_0^{N}$; the space of rays is a differential manifold called the complex projective space, and is denoted by $\mathbb{C}\mathrm{P}^{N-1}$. As a matter of fact,  $\mathbb{C}_0^{N}$ is a $\mathbb{C}_0$ principal fibre bundle whose action
\begin{equation}
 {\mathbf{z}}^\lambda = \lambda {\mathbf{z}}, \quad {\mathbf{z}}\in\mathbb{C}_0^{N}\  {\hbox{and}}\  \lambda \in \mathbb{C}_0,
\end{equation}
provides a {\it space of orbits}, the base of the bundle, given by $\mathbb{C}\mathrm{P}^{N-1}$.
We shall denote the bundle by
\begin{equation}
\mathbb{C}^{N}_0 \! \left(\mathbb{C}\mathrm{P}^{N-1},\mathbb{C}_0 \right),
\label{one}
\end{equation}
where the base space $\mathbb{C}\mathrm{P}^{N-1}$ and the Lie group $\mathbb{C}_0$ are specifically indicated.

The Hermitian scalar product 
\begin{equation} 
\langle {\mathbf{z}} | {\mathbf{w}} \rangle = \sum_{i=1}^{N} {\overline{z}_i w_i}
\label{eq:Hermitian}
\end{equation}
among vectors of the Hilbert space $\mathbb{C}^{N}$ allows to define a Hermitian tensor which coincides with the Hermitian product on $T\mathbb{C}^{N}$. In this manner we consider the Hilbert space as a Hilbert manifold so that  $\mathbb{C}^{N}$,  $\mathbb{C}^{N}_0$ and $\mathbb{C}\mathrm{P}^{N-1}$ are all Hilbert manifolds. This twist from vector spaces to manifolds is the content of the manifold (or geometrical) approach to quantum mechanics \cite{e3m,{ce3ms},bz}.

From this geometric point of view,  $\mathbb{C}^{N}_0$ becomes a Riemann manifold carrying a symplectic structure and a related complex structure.

They may be simply described in coordinates, with $\mathbb{C}^{N}_0$ thought of as a real differential manifold. The Hermitian tensor will be, for any ${\mathbf{z}}\in\mathbb{C}^{N}$,
\begin{equation}
h = \langle \mathrm{d} {\mathbf{z}} | \mathrm{d}{\mathbf{z}}\rangle.
\label{eq:hdef}
\end{equation}
If we use an orthonormal basis $|e_1\rangle, \dots, |e_{N}\rangle$, we have
\begin{eqnarray}
|{\mathbf{z}}\rangle &=& z^j |e_j\rangle = (x^j + \rmi y^j) |e_j\rangle
\nonumber\\
|\mathrm{d}{\mathbf{z}} \rangle  &=& \mathrm{d} z^j |e_j\rangle = (\mathrm{d} x^j  + \rmi \mathrm{d} y^j)|e_j\rangle ,
\end{eqnarray}
where we use Einstein notation for vector index contractions.
By spelling out~\eqref{eq:hdef} we find
\begin{equation}
h = (\mathrm{d}x^j \otimes \mathrm{d}x^j + \mathrm{d}y^j \otimes \mathrm{d}y^j) 
+ \rmi (\mathrm{d}x^j \wedge \mathrm{d} y^j),
\end{equation}
the real part is a Riemannian structure, the imaginary part is a symplectic structure. The two structures define a complex structure
\begin{equation}
J = \mathrm{d}x^j \otimes \frac{\partial}{\partial y^j} - \mathrm{d} y^j \otimes \frac{\partial}{\partial x^j}.
\label{eq:Jdef}
\end{equation}
The infinitesimal generators of the $\mathbb{C}_0$ action, the fundamental vector fields, are
\begin{equation}
\Delta = x^j \frac{\partial}{\partial x^j} + y^j \frac{\partial}{\partial y^j},
\qquad
\Gamma =  x^j \frac{\partial}{\partial y^j} - y^j \frac{\partial}{\partial x^j}.
\end{equation}
If, according to the probabilistic interpretation, we had to consider only quantities which are ``gauge invariant'', they should be invariant under the joint action of $\Delta$ and $\Gamma$. Clearly, for this to be the case, we should modify $h$ by means of a conformal factor, namely
\begin{equation}
\tilde{h} = \frac{1}{\langle {\mathbf{z}}|{\mathbf{z}}\rangle} \langle \mathrm{d} {\mathbf{z}} | \mathrm{d}{\mathbf{z}}\rangle.
\label{eq:htildedef}
\end{equation}
The connection form of the principal bundle is easily seen to be
\begin{equation}
\mathcal{A} = \Delta\otimes \frac{1}{2} \frac{\mathrm{d} \langle{\mathbf{z}}|{\mathbf{z}}\rangle}{\langle {\mathbf{z}}|{\mathbf{z}}\rangle} 
+ \Gamma \otimes \frac{1}{2} J\left(\frac{\mathrm{d} \langle{\mathbf{z}}|{\mathbf{z}}\rangle}{\langle {\mathbf{z}}|{\mathbf{z}}\rangle} \right),
\end{equation}
indeed
\begin{equation}
\mathcal{A}(\Delta) = \Delta, \qquad \mathcal{A}(\Gamma) = \Gamma.
\end{equation}
In coordinates we have
\begin{equation}
\frac{1}{2} \frac{\mathrm{d} \langle{\mathbf{z}}|{\mathbf{z}}\rangle}{\langle {\mathbf{z}}|{\mathbf{z}}\rangle} = 
\frac{1}{2} \frac{\mathrm{d} (\|{\mathbf{x}}\|^2 + \|{\mathbf{y}}\|^2)}{\|{\mathbf{x}}\|^2 + \|{\mathbf{y}}\|^2} = \frac{{\mathbf{x}}\cdot \mathrm{d}{\mathbf{x}} + {\mathbf{y}}\cdot \mathrm{d}{\mathbf{y}}}{\|{\mathbf{x}}\|^2 + \|{\mathbf{y}}\|^2}
= \vartheta ,
\end{equation}
\begin{equation}
\frac{1}{2} J\left(\frac{\mathrm{d} \langle{\mathbf{z}}|{\mathbf{z}}\rangle}{\langle {\mathbf{z}}|{\mathbf{z}}\rangle}\right)= 
 \frac{ {\mathbf{y}}\cdot \mathrm{d}{\mathbf{x}} - {\mathbf{x}}\cdot\mathrm{d}{\mathbf{y}}}{\|{\mathbf{x}}\|^2 + \|{\mathbf{y}}\|^2} = J(\vartheta).
\end{equation}
We find that $\mathrm{d}\vartheta = 0$,
\begin{equation}
\mathrm{d}J(\vartheta) = 2  \frac{ \mathrm{d}{\mathbf{y}} \wedge \mathrm{d}{\mathbf{x}} }{\|{\mathbf{x}}\|^2 + \|{\mathbf{y}}\|^2} 
- 2 \frac{({\mathbf{x}}\cdot \mathrm{d}{\mathbf{x}} + {\mathbf{y}}\cdot \mathrm{d}{\mathbf{y}}) \wedge ({\mathbf{y}}\cdot \mathrm{d}{\mathbf{x}} - {\mathbf{x}}\cdot \mathrm{d}{\mathbf{y}}) }{(\|{\mathbf{x}}\|^2 + \|{\mathbf{y}}\|^2)^2},
\end{equation}
which gives the curvature.

If we think of $\tilde{h}$ in the spirit of the Kaluza-Klein theory it is clear that the pull-back of the metric tensor on $\mathbb{C}\mathrm{P}^{N-1}$ should be
\begin{equation}
 \frac{1}{\langle {\mathbf{z}}|{\mathbf{z}}\rangle} \langle \mathrm{d} {\mathbf{z}} | \mathrm{d}{\mathbf{z}}\rangle
-  \frac{\langle \mathrm{d} {\mathbf{z}} | {\mathbf{z}} \rangle\langle{\mathbf{z}}| \mathrm{d}{\mathbf{z}}\rangle} {\langle {\mathbf{z}}|{\mathbf{z}}\rangle ^2},
\label{eq:htildepb}
\end{equation}
the second term being associated with the connection form, i.e.\
\begin{equation}
  \frac{\langle \mathrm{d} {\mathbf{z}} | {\mathbf{z}} \rangle \otimes \langle{\mathbf{z}} | \mathrm{d}{\mathbf{z}}\rangle} {\langle {\mathbf{z}}|{\mathbf{z}}\rangle ^2}
  = \vartheta \otimes \vartheta + J(\vartheta)\otimes J(\vartheta) + i  \vartheta \wedge J(\vartheta).
\end{equation}
 This can be easily computed in coordinates, or in intrinsic terms by using the properties of $J$, $J^2= - \mathbb{I}$.

It is possible also to consider the action of $\Delta$ and $\Gamma$ separately so that we identify the quotient of $\mathbb{C}^{N}_0$ under dilations to be represented by the unit sphere 
\begin{equation}
S^{2N -1} = \{{\mathbf{z}}\in\mathbb{C}^{N}\, :\, \|{\mathbf{z}}\|=1 \}
\end{equation}
and $\Gamma$ acts on the unit sphere $S^{2N-1}$ to define a $\mathrm{U}(1)$-principal bundle 
\begin{equation}
{S^{2N-1}}\!\left(\mathbb{C}\mathrm{P}^{N-1},\mathrm{U}(1)\right).
\label{three1}
\end{equation}
Now the connection one-form will simply be
\begin{equation}
\mathcal{A} = -({\mathbf{x}}\cdot\mathrm{d}{\mathbf{y}} - {\mathbf{y}}\cdot\mathrm{d}{\mathbf{x}} ), \qquad \|{\mathbf{x}}\|^2 + \|{\mathbf{y}}\|^2 =1.
\end{equation}
The symplectic structure on $\mathbb{C}\mathrm{P}^{N-1}$ represents the curvature of our connection. The second homotopy group of  the projective space  ${\mathbb{C}\mathrm{P}^{N-1}} $ is $\mathbb{Z}$, i.e. $\pi_2(\mathbb{C}\mathrm{P}^{N-1})=\mathbb{Z}$.
The first Chern number of the bundle~(\ref{one}) restricted to any non-contractible compact submanifold  $\Sigma_2\subset{\mathbb{C}\mathrm{P}^{N-1}}$ is non-trivial, i.e.
  \begin{equation}
c_1=\frac1{2 \pi}\int_{\Sigma_2} F \neq 0.
\label{chern}
\end{equation}
Moreover, 
 \begin{equation}
 \frac1{(2 \pi)^N}\int_{\mathbb{C}\mathrm{P}^{N-1}} F^{\wedge N} \neq 0,
\label{chern2}
\end{equation}
also showing that the bundle (\ref{three1}) is non-trivial.

\subsection{Uhlmann phase}

If we consider the projection from the unit sphere $S^{2N-1}=\{{\mathbf{z}}\in\mathbb{C}^{N}\, :\, \|{\mathbf{z}}\|=1 \}$ to the complex projective space, it is possible to define
\begin{eqnarray}
\pi: \mathbb{C}^{N}_0 &\to& \mathbb{C}\mathrm{P}^{N-1} 
\nonumber\\
|{\mathbf{z}}\rangle &\mapsto & \frac{|{\mathbf{z}}\rangle\langle{\mathbf{z}}|}{\langle {\mathbf{z}} | {\mathbf{z}}\rangle} = \rho_{\mathbf{z}}
\end{eqnarray}
i.e., pure states are represented by rank-one projectors.

This representation is an embedding of the complex projective space in the real vector space of Hermitian operators. This point of view is quite convenient because Hermitian matrices are isomorphic to the vector space of the Lie algebra $\mathfrak{u}(N)$  of the unitary group $\mathrm{U}(N)$. 

We can briefly review the previous arguments by means of this identification. Remember that $\mathbb{C}\mathrm{P}^{N-1} $ 
is  the manifold of rays of $\mathbb{C}^N$. At each point $\rho_{\mathbf{z}}\in\mathbb{C}\mathrm{P}^{N-1}$,
 the vectors in the tangent space $T_{\rho_{\mathbf{z}}}\mathbb{C}\mathrm{P}^{N-1}$ arise from
\begin{equation}
\frac{\mathrm{d}}{\mathrm{d} t} \frac{U(t) |{\mathbf{z}}\rangle\langle {\mathbf{z}}| U(t)^\dag}{\langle{\mathbf{z}}|{\mathbf{z}}\rangle}\Big\vert_{t=0} = -\rmi [K,{\rho_{\mathbf{z}}}] = X,
\qquad K=K^\dag, \qquad X\in T_{\rho_{\mathbf{z}}}\mathbb{C}\mathrm{P}^{N-1},
\end{equation}
where $U(t) = \exp(-\rmi t K)$ is the unitary group generated by $K$. Thus $K$ is determined by $X$ up to a matrix commuting with ${\rho_{\mathbf{z}}}$. However, this ambiguity will not affect the definition of the symplectic structure we are going to give.

If  ${\rho_{\mathbf{z}}} = |{\mathbf{z}}\rangle\langle \mathbf{z}|$ and ${\mathbf{w}}$ is a vector orthogonal to ${\mathbf{z}}$, $\langle{\mathbf{z}} |{\mathbf{w}}\rangle = 0$, we may write
\begin{eqnarray}
K &=&  \rmi (|{\mathbf{w}}\rangle\langle {\mathbf{z}}| - |{\mathbf{z}}\rangle\langle {\mathbf{w}}|),
\nonumber\\
X &=& |{\mathbf{w}}\rangle\langle {\mathbf{z}}| + |{\mathbf{z}}\rangle\langle {\mathbf{w}}|.
\end{eqnarray}

The connection one-form $\mathcal{A}$ may be used to define horizontal lifts of smooth curves in $\mathbb{C}\mathrm{P}^{N-1}$. If $\gamma = \{\rho(s) \in \mathbb{C}\mathrm{P}^{N-1} \, : \, s\in[s_1,s_2], \rho(s_1)= \rho(s_2)\} \subset \mathbb{C}\mathrm{P}^{N-1}$ is a smooth parametrised closed curve in $\mathbb{C}\mathrm{P}^{N-1}$, and 
$\gamma_h = \{{\mathbf{z}}(s) \in S^{2N-1} \, : \, s\in[s_1,s_2] \} \subset S^{2N-1} $ is a horizontal lift of $\gamma$ to $S^{2N-1}$, then at each point of $\gamma_h$ we have
\begin{equation}
\mathcal{A}_{{\mathbf{z}}(s)}\bigl(\dot{{\mathbf{z}}}(s)\bigr) = -\rmi \langle {\mathbf{z}}(s) | \dot{{\mathbf{z}}}(s)\rangle = 
2 \operatorname{Im} \langle {\mathbf{z}}(s) | \dot{{\mathbf{z}}}(s)\rangle = 0.
\end{equation}
This lift $\gamma_h$ of $\gamma$   is not closed in general, as ${\mathbf{z}}(s_1)$ and ${\mathbf{z}}(s_2)$ may differ by an element in $\mathrm{U}(1)$. This is the $\mathrm{U}(1)$ holonomy group element and gives the geometric phase associated with $\gamma$:
\begin{equation}
\operatorname{arg} \langle {\mathbf{z}}(s_1)|{\mathbf{z}}(s_2)\rangle = - \int_\Sigma \mathrm{d} \mathcal{A}, \qquad \partial \Sigma = \gamma,
\end{equation}
where $\Sigma\subset\mathbb{C}\mathrm{P}^{N-1}$ is any smooth two-dimensional surface with boundary $\gamma$.
This geometric phase is the Aharonov-Anandan phase of pure quantum states.

The bundle picture emerges very simply if we notice that given a fiducial normalised vector ${\mathbf{z}}$, such that $|{\mathbf{z}}\rangle\langle{\mathbf{z}}| = {\rho_{\mathbf{z}}}$, then $S^{2N-1}$ can be identified with $\mathrm{U}(N)/\mathrm{U}(N-1)$, as $\mathrm{U}(N)$ acts transitively on all normalised vectors and $\mathrm{U}(N-1)$ is the isotropy group of ${\mathbf{z}}$. By further modding out by $\mathrm{U}(1)$, to go from ${\mathbf{z}}$ to ${\rho_{\mathbf{z}}}=|{\mathbf{z}}\rangle\langle {\mathbf{z}}|$, we get the $\mathrm{U}(1)$-bundle
\begin{equation}
\mathrm{U}(1) \to \frac{\mathrm{U}(N)}{\mathrm{U}(N-1)} \to \frac{\mathrm{U}(N)}{\mathrm{U}(N-1) \times  \mathrm{U}(1)}.
\end{equation}
What is  remarkable is that not only $\mathrm{U}(N)$ acts on rank-one projectors but also its complexification $\mathrm{GL}(N,\mathbb{C})$. Indeed, for any vector ${\mathbf{z}}\in\mathbb{C}^{N}_0$, we have $T:{\mathbf{z}} \mapsto T{\mathbf{z}}$. To obtain a normalised vector we have to modify the action into a non-linear one
\begin{equation}
T:{\mathbf{z}} \mapsto \frac{T{\mathbf{z}}}{\sqrt{\langle T{\mathbf{z}}|T{\mathbf{z}}\rangle}},
\end{equation}
however this action passes to the complex projective space according to
\begin{equation}
\frac{T |{\mathbf{z}}\rangle\langle {\mathbf{z}}| T^\dag}{\langle T{\mathbf{z}}|T{\mathbf{z}}\rangle} = \frac{T\rho T^\dag}{\tr (T\rho T^\dag) }.
\end{equation}
Thus the complex projective space is also an orbit of $\mathrm{SL}(N,\mathbb{C})$. These remarks are useful when dealing with generic mixed states, not only pure states.

\subsection{A bundle picture  for the Uhlmann phase of mixed states}

In finite dimensions there are a few remarkable ``coincidences''. Given the complex Hilbert space $\mathbb{C}^{N}$,  the space $B(\mathbb{C}^{N})$ of linear operators is isomorphic to $\mathbb{C}^{N}\otimes\mathbb{C}^{N\ast}$ and may be considered itself a Hilbert space. $B(\mathbb{C}^{N})$ is a $C^*$-algebra, it carries a $*$-involution $A\mapsto A^\dag$. Every element $M\in B(\mathbb{C}^{N})$ may be written uniquely as $M = A + \rmi B$ , where $A=A^\dag$ and $B=B^\dag$ are Hermitian operators.
$B(\mathbb{C}^{N})$ has a Lie algebra structure and corresponds to the Lie algebra $\mathfrak{gl}(N,\mathbb{C})$ of $\mathrm{GL}(N,\mathbb{C})$, the complexification of $\mathrm{U}(N)$. The Hermitian scalar product on $B(\mathbb{C}^{N})$ given by the Hilbert-Schmidt product
\begin{equation}
\langle M_1 | M_2 \rangle =\tr ( M_1^\dagger M_2 )
\label{Hilbert-Schmidt}
\end{equation}
may be split into its real and imaginary part:
$\tr ( M_1^\dagger M_2 + M_2^\dagger M_1 ) /2$ and $-\rmi \tr ( M_1^\dagger M_2 - M_2^\dagger M_1 )/2$. It turns out that $\mathfrak{gl}(N, \mathbb{C})$ as a Lie algebra is symplectomorphic to $T^*(\mathfrak{u}(N))$, i.e.\  $\mathrm{GL}(N, \mathbb{C})$ is symplectomorphic to $T^*(\mathrm{U}(N))$.

These various ``coincidences'' have been exploited to consider $B(\mathbb{C}^{N})$ as a bundle space whose group is the unitary group and the base manifold is the space of mixed states.
The construction goes along the following lines. We consider the projection 
\begin{eqnarray}
\pi: B(\mathbb{C}^{N}) &\to& H_+
\nonumber\\
M&\mapsto& M M^\dag,
\end{eqnarray}
where $H_+$ is the space of Hermitian positive operators. Each fibre is diffeomorphic to $\mathrm{U}(N)$, indeed $M U$ and $M$ give rise to the same positive operator
\begin{equation}
 M U (M U)^\dag=  M M^\dag, \qquad \text{for all} \quad U \in \mathrm{U}(N).
\end{equation}
It may be convenient to consider also the projection $M\mapsto M^\dag M$, where the fibre will be generated by the left action of the unitary group $M\mapsto U M$, so that $M\mapsto M^\dag  M= (UM)^\dag U M$.

If we ``normalise'' our projection, say
\begin{equation}
\tilde{\pi}: M\mapsto \frac{M M^\dag}{\tr(M M^\dag)},
\end{equation}
the image of this projection does coincide with the space of all mixed states. Having defined the normalised projection by means of the right action, it follows that the left action of $\mathrm{GL}(N,\mathbb{C})$ passes to the quotient. We have
\begin{equation}
T\mapsto T M, \qquad \tilde{\pi}: TM \mapsto \frac{TM M^\dag T^\dag}{\tr(TM M^\dag T^\dag)}.
\end{equation}
By introducing the polar decomposition $M = \sqrt{\rho} U$ we find
\begin{equation}
M M^\dag = \sqrt{\rho} U U^\dag \sqrt{\rho} = \rho.
\end{equation}
The orbits generated by the action of $\mathrm{GL}(N, \mathbb{C})$, say
\begin{equation}
\frac{T\rho T^\dag}{\tr(T\rho T^\dag)},
\end{equation}
decompose the space of mixed states into strata, according to the rank of $\rho$. Therefore, the total space is partitioned into $N$-strata of orbits, the one corresponding to rank-one operators $M$ will give the complex-projective space of pure states we have considered earlier. Except for the rank-one orbit, the other orbits are not symplectic manifolds and turn out to be the union of symplectic orbits with changing dimensions. This fact is related to the circumstance that symplectic orbits are associated with the spectrum of the mixed state, while the orbits of $\mathrm{GL}(N, \mathbb{C})$ are associated with the rank of the state.

Taking into account the normalisation, it is immediate to notice that the central subgroup of $\mathrm{GL}(N, \mathbb{C})$, generated by $\lambda\mathbb{I}$, with $\lambda \in \mathbb{C}_0$, acts trivially, therefore the orbits are actually orbits of $\mathrm{SL}(N, \mathbb{C})$.

Let us now restrict ourselves to mixed states of maximal rank. They span the main stratum, a dense subset, $\mathcal{D}_0\in\mathcal{D}$ of the space of all mixed states $\mathcal{D}$. In order to define a connection which generalizes the Aharonov-Anandan connections  we need to consider a principal $U(N)$-bundle structure over $\mathcal{D}_0$ \cite{AU1,AU2,AU3,AU4}.

Let us consider the submanifold  $GL_0(N, \mathbb{C})$ of $GL(N, \mathbb{C})$  given by the matrices of unit Hilbert-Schmidt norm, i.e.
$GL_0(N, \mathbb{C})=\{ M\in GL(N, \mathbb{C}) \, : \,  \langle M | M \rangle =1\}$.

The right action of $U(N)$ on    $GL_0(N, \mathbb{C})$: $A\mapsto AU$ defines a principal fibre bundle over the 
space of mixed states of maximal rank ${\mathcal{D}}_0$,
 \begin{equation}
GL_0(N, \mathbb{C})\bigl({\mathcal{D}}_0,\mathrm{U}(N)\bigr).
\label{onec}
\end{equation}

Following  the definition of the metric in the case of pure states we can define a metric in  $GL_0(N, \mathbb{C})$ by
\begin{equation} 
g(M_1,M_2) =  \operatorname{Re} \langle M_1 | M_2 \rangle,
\label{onehalf}
\end{equation}
associated with the Hilbert-Schmidt inner product (\ref{Hilbert-Schmidt}).
This metric can be related to the Bures metric defined on the same space~\cite{Bures,CJ}.

As in the case of pure states, the Riemannian metric structure~(\ref{onehalf}) of $GL_0(N, \mathbb{C})$    
 induces a connection on the bundle~(\ref{onec}) given by the distribution
of horizontal spaces of $T_M GL_0(N, \mathbb{C})$ which are orthogonal to the $\mathrm{U}(N)$ fibres. This connection is 
the Uhlmann connection~\cite{AU1,AU2,AU3,AU4}. The connection is defined by the one-form $\mathcal{A}_U$ with values on the Lie algebra
$\mathfrak{u}(N)$ of  $\mathrm{U}(N)$ which vanish on the horizontal subspaces of $T_M GL_0(N, \mathbb{C})$.  The explicit form of the connection is more involved than that of the Aharonov-Anandan connection for pure states, although can be derived from
a quite simple analysis.

The space of horizontal vectors $T_M GL_0(N, \mathbb{C})$ is given by all vectors $X$ which satisfy
\begin{equation}
X^\dagger M - M^\dagger X=0.
\label{three}
\end{equation}
This is so, because any  vector tangent to the fibres is of the form $\rmi M \phi$ where $\phi$ is any  $N\times N$ Hermitian matrix,
and since $g(X, \rmi M\phi)=0$, it follows that $\operatorname{Re} \tr ( \rmi  X^\dagger M \phi) =0$. Since the equation holds for any Hermitian matrix $\phi$, $X^\dagger M$ must be Hermitian, which implies~(\ref{three}).

In the same way it can be shown that the  vertical vectors of $T_M GL_0(N, \mathbb{C})$ which are tangent to the gauge fibres are of the form  
$Y_\phi=\rmi M\phi$ and  can be identified with the solutions of 
\begin{equation}
Y_\phi^\dagger M + M^\dagger Y_\phi=0.
\label{threeb}
\end{equation}

Therefore, the one form $\mathcal{A}_U$ characterising Uhlmann connection has to satisfy the two conditions
 \begin{equation}
\mathcal{A}_U (X)=0 \quad \mathrm{and } \quad \mathcal{A}_U(Y_\phi)=\rmi\phi.
\label{fourth}
\end{equation}
This implies that 
 \begin{equation}
\mathcal{A}_U M^\dagger M+ M^\dagger M \mathcal{A}_U=M^\dagger \mathrm{d} M -(\rmd M^\dagger) M.
\label{fourthb}
\end{equation}
Notice that from this relation it follows that $\mathcal{A}_U$ take values in $\mathfrak{u}(N)$ and vanish for horizontal tangent vectors.

This implicit formula can be made more explicit if we consider an orthonormal basis 
$\{|e_j\rangle, \, j=1,2,\dots ,N\}$ of $\mathbb{C}^N$
which diagonalises the positive definite matrix~$M^\dagger M$,
 \begin{equation}
 M^\dagger  M |e_j \rangle= c_j|e_j \rangle.
\label{fourtha}
\end{equation}
 In such a case
 \begin{equation}
\langle e_j |\mathcal{A}_U M^\dagger M|e_k \rangle+ \langle e_j | M^\dagger M \mathcal{A}_U|e_k \rangle=\langle e_j |M^\dagger \rmd M|e_k \rangle -\langle e_j| (\rmd M^\dagger) M|e_k \rangle
\label{fourthbb}
\end{equation}
which implies, by setting $M = U \sqrt{M^\dagger M}$,
 \begin{eqnarray}
 (c_k+ c_j) \langle e_j | \mathcal{A}_U|e_k \rangle 
&=& \langle e_j | M^\dagger \rmd M|e_k \rangle -\langle e_j | (\rmd M^\dagger) M|e_k \rangle
\nonumber\\
&=& \langle e_j| \bigl[ \, \sqrt{M^\dagger M}, \rmd \sqrt{M^\dagger M} \,\bigr]|e_k \rangle
+2 \sqrt{c_j c_k} \langle e_j| U^\dagger \rmd U |e_k \rangle,
\end{eqnarray}
whence 
 \begin{equation}
\langle e_j | \mathcal{A}_U|e_k \rangle=\frac{\langle e_j |  \bigl[\, \sqrt{M^\dagger M}, \rmd \sqrt{M^\dagger M} \, \bigr] |e_k \rangle}{c_k+c_j}.
\label{fourthbbb}
\end{equation}
The holonomy of this connection is the Uhlmann phase~\cite{AU1,AU2}.

\subsection{Remark}

The bundle~(\ref{onec}) we have constructed   over the space of  maximal rank mixed states is trivial, unlike the one we have constructed for pure states.

Indeed the base manifold is contractible being diffeomorphic to a vector space (this may be seen from the diffeomorphism $\mathrm{GL}(N, \mathbb{C}) \simeq T^* \mathrm{U}(N)$). A direct proof follows from the polar decomposition $M = \sqrt{\rho} U = \sqrt{M M^\dag} U$, which allows to define a global section $\sigma: \mathcal{D}_0 \to S^{2N-1}\subset \mathrm{GL}_0(N, \mathbb{C})$ given by $\sigma (\rho) = M_\rho = \sqrt{\rho}$ \cite{bd}.

Thus, all the characteristic classes of the bundle vanish. In particular,
for any $n=1,2, \cdots, \left[(N^2-1)/{2}\right]$,
where the brackets denotes the integer part, one gets
 \begin{equation}
 c_n(\mathcal{A}_U)=\frac1{(2 \pi)^{2n}}\int_{\Sigma^{2n}}\tr \left\{ F(\mathcal{A}_U)^{\wedge 2n} \right\} =0,
\label{chern-uhlmann1}
\end{equation}
for any closed $2n$-dimensional submanifold ${\Sigma^{2n}}$ of $\mathcal{D}_0$.

This result is quite remarkable  because it is in contrast with what happens for pure states, where the connection that generates the  Aharanov-Ananadan phase is topologically non-trivial, and the corresponding Chern class does not vanish, whereas
for mixed states the connection which generates the Uhlmann phase has vanishing Chern classes. 

Among the states which are of maximal rank there are {\it thermal states}:
\begin{equation}
\rho_T = \frac{e^{-H/T}}{\tr(e^{-H/T})}.
\label{thermal}
\end{equation}
Now,  thermal states 
converge  in the zero temperature limit  to a pure state, provided that the
 Hamiltonian $H=H^\dagger$ has a non-degenerated ground state $|0\rangle$, i.e.
\begin{equation}
\lim_{T\to 0} \rho_T = |0\rangle\langle 0|.
\end{equation}

This leads to a surprising phenomenon: the emergence of topological order in the zero temperature limit of thermal states.
How the triviality of the Uhlmann phase topology for any finite temperature can lead to the non-trivial  Aharonov-Anandan  topology 
 in the  zero temperature  limit~\cite{RVMD,RVMD1,RVMD2,RVMD3,HGC}? This is a well posed problem which deserves to be understood.

In order to find a framework where the non-trivial topology can play a role in the dynamics of mixed states
it is convenient to exclude from the space of mixed states the state of maximal entropy which correspond to maximal disorder
\begin{equation}
\rho_\ast= \frac{1}{N} \mathbb{I}
\label{disorder}
\end{equation}
This extra requirement might be physically motivated by the fact that  
Gibbs thermal  states $\rho_T$~(\ref{thermal}) of non-degenerated Hamiltonians are only
maximally disordered at infinity temperature, i.e.  $\rho_T$~(\ref{thermal})
 belongs to the main strata of mixed states $\mathcal{D}_0$ for any finite temperature $T$. 
Therefore, excluding such a singular states will be natural for generic thermal systems.

In such a case it is trivial to see that the corresponding space of physical states $\mathcal{D}_\ast=\mathcal{D}_0\backslash\{\rho_\ast\}$
acquires a non-trivial topology. In fact $\mathcal{D}_\ast$ becomes homeomorphic to  $S^{N^2-2}\times (0,1)$
which inherits the non-trivial topology of the sphere  $S^{2N-1}$. However, the restriction of the bundle~(\ref{onec}) to $\mathcal{D}_\ast$ 
\begin{equation}
GL_\ast(N, \mathbb{C}) \bigl({\mathcal{D}}_\ast,\mathrm{U}(N)\bigr),
\label{fourb}
\end{equation}
where $GL_\ast(N, \mathbb{C})=\{M\in  GL_0(N, \mathbb{C}) \, : \, \det \, M=1/{\sqrt{N}}\}$ 
is again a  trivial bundle because the section  $\sigma: {\mathcal{D}}_\ast\to GL_\ast(N, \mathbb{C})$ given by $\sigma(\rho)=\sqrt{\rho}$ is a 
also a global section in the new framework too.

For such a reason there  have been many attempts to define new topological invariants which extend the topological order to thermal states at finite temperatures~\cite{RVMD,RVMD1,RVMD2,RVMD3,HGC,Mera1,Mera2,Mera4,Car1,Car2,Car3}. For instance, one proposal is to define the modified Chen classes by weighting then  with the  density matrix, e.g. 
 \begin{equation}
 c_n(\mathcal{A}_U)=\frac1{(2 \pi)^{2n}}\int_{\Sigma^{2n}}\tr \Bigl\{M M^\dagger F(\mathcal{A}_U)^{\wedge 2n} \Bigr\}. 
\label{chern-uhlmann}
\end{equation}
Unfortunately  this approach is not stable under small perturbations.

If  we have a family of non-maximally-disordered mixed states periodically depending on $N^2-2$  parameters $\rho(\epsilon_1,
\epsilon_2,\cdots, \epsilon_{N^2-2})$ there is a new possibility for the introduction of new  topological invariants
by means of  the winding number 
\begin{equation}
\nu=\frac1{(2\pi)^{N^2-2}} \int_{\bT^{N^2-2}} \tr  \Bigl\{ (\rho^{-1} \rmd\rho)^{\wedge N^2-2} \Bigr\}
\label{winding}
\end{equation}
of
the map $\rho: {\bT^{N^2-2}}\to \mathcal{D}_0$.
The winding number of the  map $\rho$ is a topological invariant which is stable under smooth perturbations. In particular, it might survive in the zero temperature limit. This opens a new perspective in the analysis of topological matter at finite temperature, extending the standard theory which has been formulated at zero temperature.

\section{Topology and Floquet-Bloch theorems}

In order to analyse  the extension of recent analyses of topological matter to finite temperature
and open dynamics let us consider  the example of an  electron in a perfect crystal.

Electrons moving in  the 
periodic  potential of a  perfect crystal split their  Hilbert space 
of  quantum states on a bundle of Floquet-Bloch states over 
a $d$-dimensional torus $\bT^d$. If we assume a cubic symmetry the
translation symmetry group $\bZ^d$ is generated by 
the lattice  of space translations $T_j=\rme^{\rmi a p_j }, j=1,2,\dots,d$
where $a$ is the crystal cell period and ${p_j}=-\rmi \partial_j$
the momentum operator.

The symmetry of the crystal implies that the Hamiltonian
is  invariant under lattice translations
\begin{equation}
T_jH= H T_j, \quad \mathrm{or}\quad  T_jHT_j^\dagger =H.
\label{eq:Hinvariance}
\end{equation}
The Hilbert space can be decomposed as a direct sum (in fact, a direct integral) of irreducible
representations of  $\bZ^d$

\begin{equation}
\mathcal{H}=\bigoplus_{\epsilon\in \bT^d} \mathcal{H}_\epsilon
\label{eq:1}
\end{equation}
where $\epsilon\in \bT^d$ is the label of the irreducible representation of $\bZ^d$,
i.e.
\begin{equation}
T_j{\mathbf{z}}_\epsilon(x)= \rme^{\rmi a \epsilon_j} {\mathbf{z}}_\epsilon (x) \qquad 0\leq \epsilon_j\leq \frac{2 \pi}{a}
\label{eq:2}
\end{equation}
which defines the Brillouin zone of the crystal. 
Now, since the periodic Hamiltonian~$H$ of the system commutes with the symmetry
group, it can be decomposed into diagonal blocks labeled by the irreducible representations
of $ \bZ^d$,
\begin{equation}
H= \bigoplus_{\epsilon\in \bT^d} H_\epsilon.
\label{sec:3}
\end{equation}
More precisely, the Floquet-Bloch decomposition defines a bundle over $\bT^d$
whose fibres are the Hilbert spaces $\mathcal{H}_\epsilon$ defined by the states satisfying the 
 conditions~(\ref{eq:2}), i.e.
 \begin{equation}
{\mathbf{z}}_\epsilon(x + a e_j)
= \rme^{\rmi a \epsilon_j} {\mathbf{z}}_\epsilon (x),
\label{sec:4}
\end{equation}
where $e_j$ is the $j$-th vector of the canonical basis of $\mathbb{R}^d$. 
If there are $N$ bands the Hilbert spaces ${\mathcal{H}}_\epsilon$ are 
$N$-dimensional  ${\mathcal{H}}_\epsilon\simeq{\mathbb{C}}^N$ , and the bundle is a rank-$N$ bundle $E(\bT^d,{\mathbb{C}}^N)$. Pure states of the solid bands are sections of
such a bundle. 

Now if the bundle is non-trivial as in the two-dimensional ($d=2$) integer Hall effect, its  non-trivial topology gives rise to interesting conducting/insulating properties characterised by topological invariants~\cite{TKNN, smt}. These phenomena can be associated with the appearance of non-trivial phases in periodic cycles of pure band states. The phases are not pure Aharanov-Anandan phases because they are twisted by the dynamics induced by  the connection $\mathcal{A}$ of the bundle $E$. Using the Fourier-Mukai transform~\cite{mukaia,mukaib} we can associate with $\mathcal{A}$ another connection $\hat{\mathcal{A}}$  in the dual bundle $\hat{E}(\hat{\bT}^2,{\mathbb{C}}^k )$ with rank $k$, the first Chern number $k=c_1(E)$ of $E$, and with $c_1(\hat{E})=N$~\cite{mukaia,mukaib,smt}. This duality transformation is the  source of quantisation of the Hall conductivity~\cite{TKNN}.

However,  these topological arguments cannot be extended to the case of mixed systems because if we consider mixed states which are invariant under the $\bZ^2$ lattice symmetry,
\begin{equation}
T_j\rho=\rho   T_j,
\end{equation}
they   can be decomposed as a sum of mixed states  on the first Brillouin zone $\bT^2$
\begin{equation}\rho=
\bigoplus_{\epsilon\in \bT^2} \rho_\epsilon,\
\label{eq:rhofibre}
\end{equation}
or in other terms as a section in the bundle of ${\mathcal{E}}(\bT^2, H_+)$ associated with $E$ by the adjoint representation of 
$\mathrm{U}(N)$, where $H_+$ denotes the space of non-negative Hermitian operators in ${\mathbb{C}}^N$. In such a case there is no phase associated with $\rho$ in  periodic cycles of $\bT^2$. However, the global map given by the section $\rho$ can have a non-trivial winding number given by
\begin{equation}
\nu_2=\frac1{2\pi} \int_{\bT^{2}} \tr  \bigl\{ \rho_\epsilon^{-1} \rmd\rho_\epsilon\wedge\rho_\epsilon^{-1} \rmd\rho_\epsilon \bigr\}
\label{winding2}
\end{equation}
where we assumed that $\rho_\epsilon$ is not pure for any $\epsilon$ nor maximally disordered, i.e. $\det\rho_\epsilon\neq 0$
and $\rho_\epsilon\neq \alpha\mathbb{I}$.

In such a case we can associate a non-trivial topology with the mixed states  with $\nu_2\neq0$, but the physical  effects of such
a property are unclear because they cannot be related to the quantisation of Hall conductivity. In fact there exist topologically non-trivial mixed states even in
absence of magnetic fields, i.e. when the bundle $E$ is trivial.

\section{Open systems}

The dynamics of open systems is governed by the GKLS equation~\cite{Koss, Lindblad, GKS}
\begin{equation}
\dot\rho=-i[H,\rho]+\sum_{i=1}^{N^2-1} \Bigl(K_i \rho K_i^\dagger -\frac12 \{K_i^\dagger K_i, \rho\}\Bigr),
\end{equation}
where $H=H^\dagger$ is the generator of the unitary part of the evolution, while the jump operators $K_i$ yield decoherence and dissipation.  

In the case of a particle in a periodic lattice one can generalise the analysis of the previous section, provided that, together with the Hamiltonian invariance~\eqref{eq:Hinvariance}, one also has
\begin{equation}
 T_jK_iT_j^\dagger =K_i,
\end{equation}
In such a case, all jump operators have a direct sum decomposition on the first Brillouin zone, $\epsilon \in \bT^2$,
\begin{equation}
H= \bigoplus_{\epsilon\in \bT^d} H_\epsilon, \qquad K_i= \bigoplus_{\epsilon\in \bT^d} K_{\epsilon,i}.
\end{equation}
Moreover, since
\begin{equation}
 T_j\dot \rho T_j^\dagger = \dot \rho,
\end{equation}
one gets that the block decomposition~\eqref{eq:rhofibre} is preserved and the evolution of each block is decoupled from the others, namely
\begin{equation}
\dot\rho_\epsilon=-i[H_\epsilon,\rho_\epsilon]+\sum_{i=1}^{N^2-1}\Bigl(K_{\epsilon,i} \rho_\epsilon K_{\epsilon,i}^\dagger -\frac{1}{2} \{K_{\epsilon,i}^\dagger K_{\epsilon,i}, \rho_\epsilon\}\Bigr),
\end{equation}
for all $\epsilon \in \bT^2$.

Now, one can show that the winding number~(\ref{winding2}) is invariant under the evolution of the open system,
because 
\begin{equation}
\nu_2(\dot{\rho}_\epsilon)=0,
\end{equation}
whenever for any value of $\epsilon$ the time evolution does not drive the system into the maximally disorder state $\rho_\ast$ 
or into lower rank  states.

The topological stability of the open system is essentially due to the continuity of the
dynamical evolution driven by the GKLS equation~\cite{egc2}.
The analysis and classification of topological transitions where the winding number jumps is an open interesting problem
which deserves further study.

From the topological viewpoint the  behaviour of the dynamics of open systems  is  richer than that driven by  the adiabatic evolution of thermal systems, where such a transition only occurs in the extreme limits of zero and infinity temperature.

\begin{acknowledgement}
MA thanks to M.A.~Martin-Delgado for stimulating discussions.
GM would like to thank the support provided by the Santander/UC3M Excellence Chair Programme 2019/2020.
This work  is partially supported by Spanish MINECO/FEDER
grant FPA2015-65745-P and DGA-FSE grant E21-17R, and COST
action program QSPACE-MP1405, by Istituto Nazionale di Fisica Nucleare (INFN) through the project ``QUANTUM'', and  by the Italian National Group of Mathematical Physics (GNFM-INdAM).

\end{acknowledgement}

\end{document}